\documentclass[10pt,journal]{IEEEtran}

\usepackage{outline}
\usepackage{pmgraph}
\usepackage[normalem]{ulem}
\usepackage[utf8]{inputenc}
\usepackage{amssymb}
\usepackage{hyperref}
\usepackage{amsmath}
\usepackage{graphicx}
\usepackage{times}
\usepackage{xcolor}
\usepackage{xspace}
\usepackage{cite}
\usepackage{bm} 
\usepackage{url}

\usepackage{float}

\usepackage{epstopdf}
\AppendGraphicsExtensions{.tiff}
\graphicspath{{fig/}} 

\usepackage{epsfig}
\usepackage{tikz}
\usetikzlibrary{spy}
\usepackage{algpseudocode}
\usepackage{algorithm}
\usepackage{mathrsfs}

\long\def\comment#1{} 

\newcommand{\xmath}[1] {\ensuremath{#1}\xspace}
\newcommand{\blmath}[1] {\xmath{\bm{#1}}}

\newcommand{\Xb}{{\blmath X}}
\newcommand{\Yb}{{\blmath Y}}
\newcommand{\Zb}{{\blmath Z}}

\newcommand{\vb}{{\blmath v}}
\newcommand{\wb}{{\blmath w}}

\newcommand{\yb}{{\blmath y}}

\newcommand{\Nc}{\mathcal{N}}

\newcommand{\Tc}{\mathcal{T}}

\newcommand{\Thetab}{{\boldsymbol {\Theta}}}

\newcommand{\Rd}{{\mathbb R}}

\newcommand{\beq}{\begin{equation}}
\newcommand{\eeq}{\end{equation}}
\newcommand{\beqa}{\begin{eqnarray}}
\newcommand{\eeqa}{\end{eqnarray}}

\begin{document}
\title{Phase Aberration Robust Beamformer for Planewave US  Using Self-Supervised Learning}
\author{Shujaat Khan, Jaeyoung Huh,~
        and~Jong~Chul~Ye,~\IEEEmembership{Fellow,~IEEE}
\thanks{Shujaat Khan and Jaeyoung Huh are with the Department of Bio and Brain Engineering, and Jong Chul Ye is with the Kim Jaechul Graduate School of AI, Korea Advanced Institute of Science and Technology (KAIST), Daejeon 34141, Republic of Korea (e-mail:\{shujaat,woori93,jong.ye\}@kaist.ac.kr). 
		}}

\maketitle

\begin{abstract}
Ultrasound (US)  is widely used for clinical imaging applications thanks to its real-time and non-invasive nature. However, its lesion detectability is often limited in many applications  due to the phase aberration artefact caused by variations in the speed of sound (SoS) within body parts. 
To address this, here we propose a novel self-supervised 3D CNN that enables  phase aberration robust plane-wave imaging.
Instead of aiming at  estimating the SoS distribution as in conventional methods, our approach is unique in that the network is trained in a self-supervised manner to robustly generate a high-quality image from various phase aberrated images by modeling the variation in the speed of sound as stochastic. Experimental results using real measurements from tissue-mimicking phantom and \textit{in vivo} scans confirmed that the proposed method can significantly reduce the phase aberration artifacts and improve the visual quality of deep scans.
\end{abstract}

\begin{IEEEkeywords}
Ultrasound imaging, phase aberration, coherent planewave compounding, 3D CNN, speed-of-sound, deep learning, beamforming.
\end{IEEEkeywords}

\IEEEpeerreviewmaketitle

\section{Introduction}
\label{sec:introduction}
Ultrasound (US) imaging is based on the time-reversal principle, in which individual channel RF measurements are back-propagated and accumulated to form an image after applying specific delays.
Although conventional B-mode focused imaging can produce high-quality images, it could increase acquisition time in a variety of clinical applications that require a large number of scan lines \cite{Imbault2017,Deffieux2018,Demene2019}. 

 In order to accelerate the scanning process, planewave imaging (PWI) schemes are often used to acquire a complete image in a single transmit/receive event. Unfortunately, the images generated using individual transmit events are of low quality, so  a single high quality image is formed with the help of coherent compounding (CPC)  of multiple low quality images acquired at different angles \cite{montaldo2009coherent}.  
However, resulting images from conventional coherent planewave compounding methods are highly susceptible to the measurement loss. In addition, the spreading of the pulse-echo system response  from speed-of-sound (SOS) errors have a further negative impact on the visualization of structures, especially in deep scans\cite{anderson2000impact}.

\begin{figure*}[!hbt]
	\centerline{\includegraphics[width=12cm]{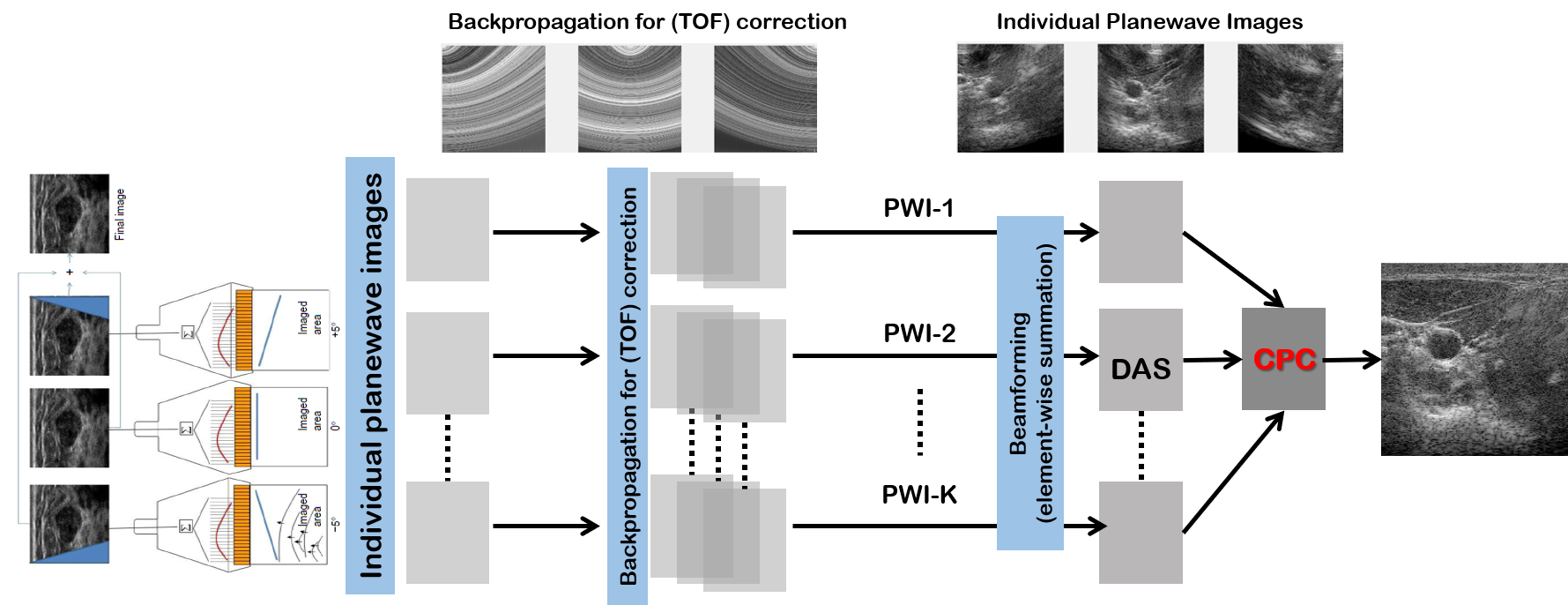}}
	\caption{Overview of the coherent planewave compounding scheme.}
	\label{fig:PWI_overview}	
\end{figure*} 

\begin{figure*}[!hbt]
	\centerline{\includegraphics[width=0.7\textwidth]{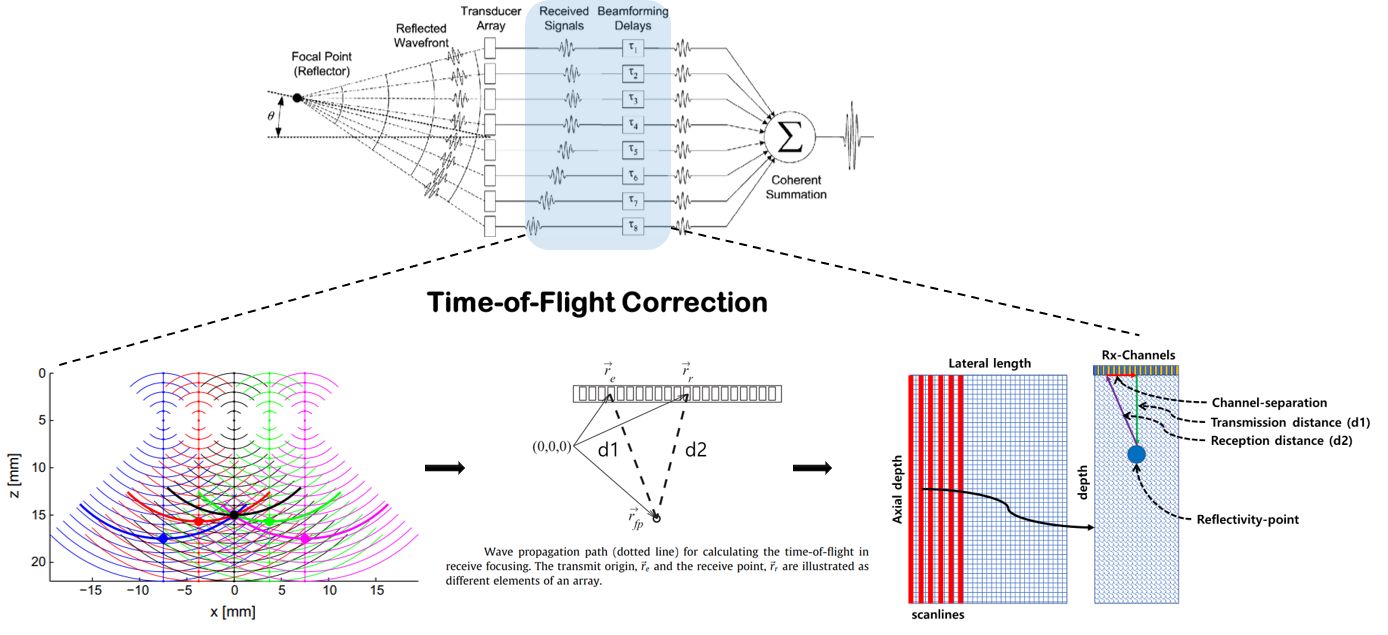}}
	\vspace*{-0.5cm}
	\caption{Overview of time-of-flight (TOF) corrections step in delay and sum beamforming pipeline.}
	\label{fig:TOF_and_SoS}	
\end{figure*}

Specifically, Fig.~\ref{fig:PWI_overview} provides an overview of the conventional coherent plane wave compounding (CPC) and outputs of individual steps, such as scanning, time-of-flight correction, and generation of individual planewave images using delay-and-sum (DAS) beamformer. Fig.~\ref{fig:TOF_and_SoS} also explains the time-of-flight (TOF) correction step in delay and sum beamforming pipeline.
Here, the time-delay used in DAS beamformer is usually calculated by
\begin{equation}\label{eq:tof}
	t_{tof} = \frac{d}{c},
\end{equation}
where $t_{tof}$ is the total time of flight, $c$ is the estimated  SoS, $d=d_1+d_2$ is the total distance travel by a sound wave during transmission of $d_1$ and reception of $d_2$ distances. 

As it can be seen in the Fig.~\ref{fig:PWI_overview} and Fig.~\ref{fig:TOF_and_SoS},  the time-of-flight correction using \eqref{eq:tof} is the first major step in the ultrasound image formation process also known as beamforming, which heavily relies on an assumed constant SoS. Nonetheless,
 in \textit{in vivo} scans of soft tissues, the SoS ranges from approximately $1450$ m/s in fat to over $1600$ m/s in muscle\cite{goss1978comprehensive}. 
 In fact, SoS varies greatly depending upon the medium it is traveling through. The more rigid (or less compressible) the medium, the higher the SoS.  
However, as a constant value of $1540$ m/s SoS is mostly used  in commercial ultrasound, this can produce sound speed errors on the order of 5\%.

This error, often called the phase aberration, usually leads to the mismatch of return echo signals of each element in the transducer and it often results in the deterioration of contrast and lateral resolution in US images. 
 Furthermore, due to the unfocused nature of PWI,  these sound speed errors significantly reduce the contrast and spatial resolution of PWI.  Accordingly,
 these variation in SoS can result in artifacts which make the clinical diagnosis an arduous task \cite{anderson2000impact}.

The effect of sound speed mismatch has already been widely studied \cite{goss1978comprehensive, mallart1992sound,anderson2000impact}. For instance, in \cite{mallart1992sound}, Mallart et al studied the effect of sound speed fluctuations in medical ultrasound imaging and provided a comparison between various correction algorithms. In another study, the effects of SoS and the attenuation on ultrasound lateral resolution were investigated using computer simulations \cite{chen2004simulation}. Since the assumed SoS does not always correspond to a exact value, the calibration causes distortion in the wave signal which affects the lateral resolution in the US images \cite{pinton2011sources}. Therefore, the correct estimation of the SoS is of central importance in order to avoid phase aberration artifacts in US images.

Various methods for phase aberration correction have been proposed, which include coherent factor-based beamforming \cite{mallart1994adaptive, nilsen2010wiener}, time delay estimation based on the sum of absolute differences between RF samples \cite{karaman1993phase}, dynamic near-field correction \cite{li1996phase}, a method using speckle brightness as a quality factor \cite{Nock1989}, etc.
In \cite{napolitano2006sound}, an algorithm was developed which determines the sound speed using various trial sound speeds that improve the image quality by analyzing the spatial frequency content in a single B-mode frame of channel data. In \cite{shin2010estimation}, a deconvolution ultrasound modelling is utilized to estimate the average SoS. In \cite{fontanarosa2011ct}, Davide {et al} designed a computed tomography-based phase aberration correction method for ultrasound guided radiotherapy. Similar methods are proposed for the magnitude of SoS aberration corrections for guided radiotherapy of prostate and other anatomical sites \cite{fontanarosa2012magnitude,fontanarosa2013speed}. 
Other methods include model-based approaches: e.g., ultrasound attenuation measurements using a reference phantom with sound speed mismatch \cite{nam2011ultrasound}, a full correction method for spatially distributed SoS via transmit beam steering \cite{jaeger2015full}, the local SoS map estimation \cite{jakovljevic2018local,rau2019ultrasound}, etc. 

Unfortunately, most of the aforementioned phase aberration correction methods are based on simulation or model-based studies which often require stringent alignment between assumed and real measurement conditions. Furthermore the methods are based on iterative/trial-based optimization principals which are computational expensive and unsuitable for real-time ultra fast ultrasound imaging \cite{Nock1989,Shapoori2015}. Moreover, these approaches are mainly designed for focused B-mode imaging.

Recently, a few learning-based phase aberration correction methods have also been presented. The work in \cite{jeon2020deep} presents a deep learning method based on auto-encoder for the correction of aberration caused by the uniform estimation of the SoS in US imaging. Other deep CNN-based SoS calibration, estimation and reconstruction methods are presented in \cite{jush2020dnn, salehi2017precise, anas2019cnn}. A deep CNN-based approach where the phase aberrator profiles are estimated using B-mode US images for phase aberration correction is presented in \cite{sharifzadeh2020phase}. In particular, Sharifzadeh et al  \cite{sharifzadeh2020phase} designed a simulation study using convolution neural network. Their CNN-based method was trained and tested on simulation phantom data to predict aberrator profile of individual elements. However, the method is also designed for focused model imaging and requires target aberration profiles which are not available for real measurement data.  

Recently, Bendjador et al   \cite{Bendjador2020} proposed a state-of-the-art  aberration correction method specifically designed for planewave imaging. Their method is based on singular value decomposition (SVD) method. The SVD-based beamforming method shows promising results compared to CPC in reducing the phase aberration artifacts. However, the major limitation of the SVD based method is its dependency on large number of acquisition angles  and extensive computational
time \cite{Bendjador2020}.

To deal with both the measurement challenge and phase aberration correction issues,  motivated by the success of deep learning application for the improvement of the quality of PWI using limited number of measurement \cite{khan2020adaptive,khan2019deep}, 
herein we propose a method for the phase aberration correction using (limited) number of planewaves.  
One of the important contributions of this work is that in contrast to the existing deep learning approaches that try to estimate the SoS distribution, we are interested in developing a phase aberration {\em robust} beamforming method without even estimating the SoS distribution. 

This idea of designing phase-aberration robust beamforming is somewhat related to the SVD-based beamforming \cite{Bendjador2020}.
However, in contrast to the SVD-based beamforming, our method is computationally  efficient for real-time
processing. 
 Moreover,  it produces  high quality imaging  even from limited number acquisition angles  as shown similarly in our  previous study \cite{khan2020adaptive,khan2019deep}.
 
 The key idea arises from our assumption that SoS variation can be viewed as random noise around the commonly used SoS value of 1540m/s.
Specifically,  our 3D CNN network is trained in a self-supervised manner by learning the mapping
between beamformed images from  RF data with randomly perturbed SoS values to the clean image with the correct SoS value.
Thanks to our novel self-supervised training scheme,
the proposed method is able to improve the quality of ultrasound images that are corrupted with unknown phase aberration. Furthermore, it improves the quality of input image when the number of input planewaves is smaller  than the normal case. 
Therefore, we believe that this is a first  universal deep beamformer
that can directly process planewave data of different sound speeds and measurement configurations to generate high quality US images. 
%
%

The rest of the paper is organized as follows. The detailed information of the proposed method are presented in Section \ref{sec:theory}, which is followed by the experimental
method in Section \ref{sec:methods} and results and discussion in Section \ref{sec:results}. Finally, the paper is concluded in Section \ref{sec:conclusion}.

\section{Theory}\label{sec:theory}
\begin{figure*}[ht]
	\centerline{\includegraphics*[width=0.8\textwidth]{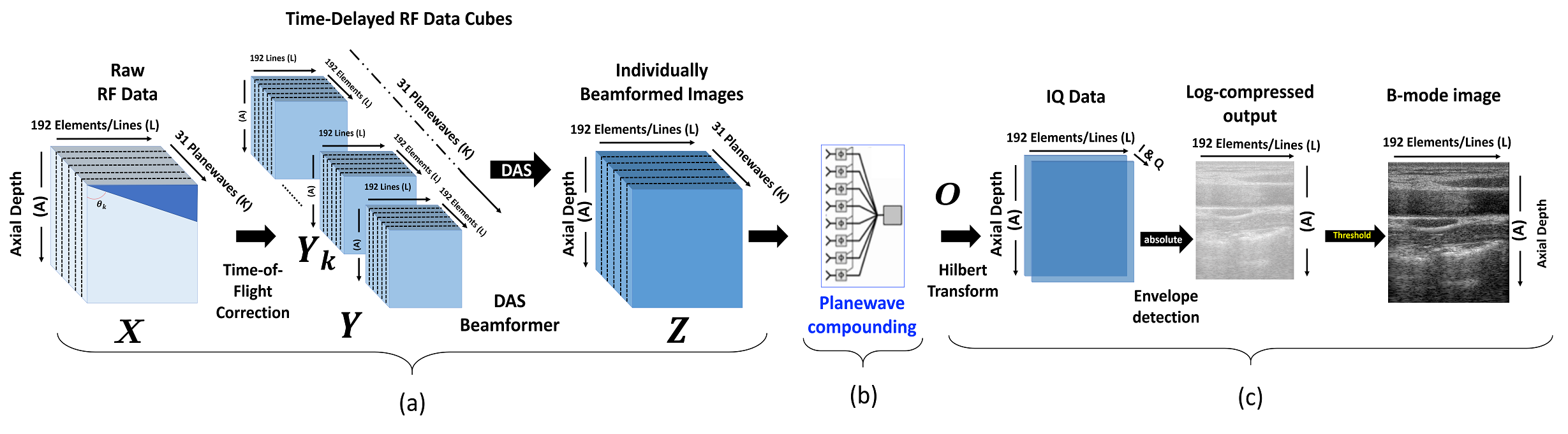}}
	\caption{Schematic of the data processing pipeline of plane wave ultrasound imaging. (a) Conversion of RAW planewave measurement data to time of flight corrected individual images using delay-and-sum (DAS) beamformer, (b) accumulation of multiple plane waves, and (c) conversion to in-phase and quadrature (IQ) data and log-compression for dynamic-range threshold.}
	\label{fig:system_block_diagram}	
\end{figure*} 

\subsection{Notation and Preliminaries}

Consider an imaging scenario shown in Fig.~\ref{fig:system_block_diagram}(a). The RF data cube $\Xb\in \Rd^{A\times L\times K}$ is represented as $$\Xb:=\left\{x^{k}_{a,l}\right\}_{a,l,k=1}^{A,L,K},$$ where  $A$, $L$ and $K$  are the total number of axial depth, lateral samples, and the plane waves, respectively; and 
$x^{k}_{a,l}$ denotes the RF data measured by the receiver channel from the $l$-th scan line at the depth index $a$ from the $k$-th planewave insonification.

The RF data of each planewave is processed using a conventional delay-and-sum (DAS) beamforming pipeline to generate time-delay corrected data cube
$\Yb_k\in \Rd^{A\times L\times L}$, which can be also represented by
$\Yb_{k}=\{\yb^k_{a,l}\}_{a,l=1}^{A,L}$ for $k=1,\cdots,K$, where
 \begin{equation}
 \yb^k_{l,a}=\begin{bmatrix} y_{l,a}^{k}[0] & y_{l,a}^{k}[1] & \cdots &y_{l,a}^{k}[L-1]\end{bmatrix}^\top \in \Rd^{L}
 \end{equation}
with
\begin{align*}
y_{a,l}^{k}[i]:=x^{k}_{a-\tau^k_{a,l}[i],l}
\end{align*}
where $^\top$ denotes the transpose, $\tau_{a,l}^k[i]$ is the time delay for the RF signal of the $k$-th planewave and the $i$-th receiver elements to obtain the $l$-th  scan line at the depth index $a$.
 
 
In the standard coherent-planewave compounding using DAS beamformer with the $K$ number of planewaves, 
the  beamformed image at the $l$-th scanline at the depth sample $a$ can be expressed as
\begin{equation}\label{eq:DAS}
{v}_{a,l} := \sum_{k=1}^{K} {z}_{a,l}^k 
=\frac{1}{L}\sum_{k=1}^{K} \wb_{a,l}^\top\yb_{l,a}^{k} 
\end{equation}
where $\wb_{a,l}$ denotes a $L$-dimensional column-vector of apodization weights
and  ${z}_{a,l}^k$ denotes the element  at the $l$-th scan line at the depth index $a$ from the $k$-th planewave insonification
form the compound tensor 
 $$\Zb:=\left\{z^{k}_{a,l}\right\}_{a,l,k=1}^{A,L,K},$$
which is processed by DAS beamformer.  

Besides apodization and compounding of planewave shown in Fig.~\ref{fig:system_block_diagram}(b), another important step in beamforming is to convert the processed data to generate the in-phase (I) and quadrature (Q) representation.
More specifically, this process is performed by Hilbert transform as shown in Fig.~\ref{fig:system_block_diagram}(c).
The filter kernel for Hilbert transform is in principle one-dimensional since it is applied along the depth direction.  The IQ representation can also be generated before beamforming and compounding step. 
Finally, as shown in Fig.~\ref{fig:system_block_diagram}(c), the detected envelopes are combined and log-compressed to generate B-mode image at particular dynamic-range threshold.

\subsection{SVD Beamformer}
\begin{figure}[ht]
	\centerline{\includegraphics*[width=0.5\textwidth]{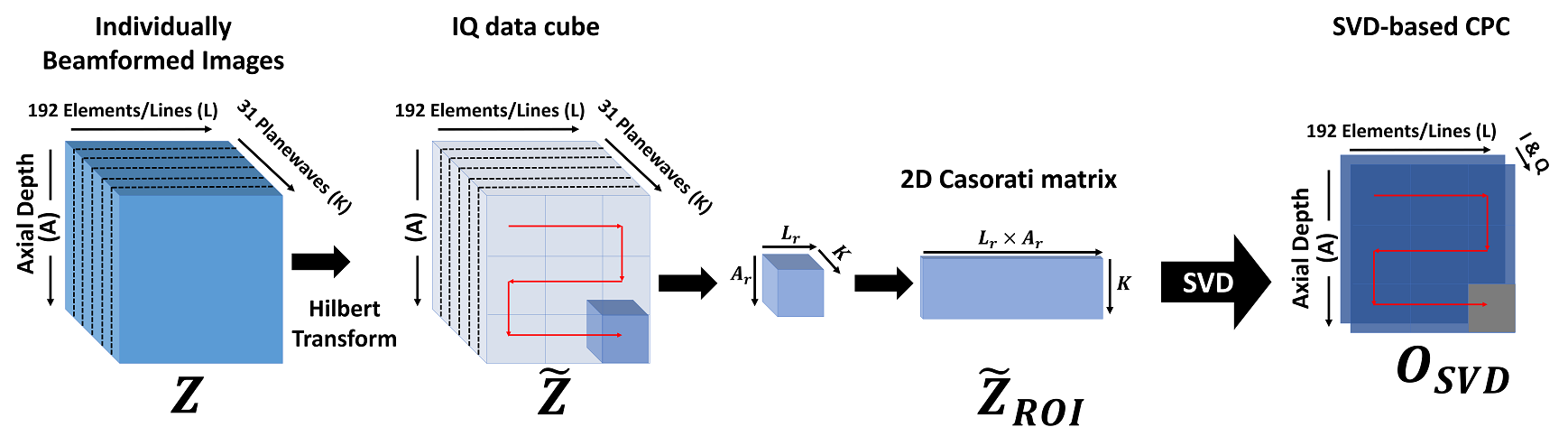}}
	\caption{The schematic of SVD Beamformer for local angular coherence-based compounding of planewave Images. }
	\label{fig:system_block_diagram_SVD}	
\end{figure} 

Fig.~\ref{fig:system_block_diagram_SVD} illustrates SVD-based CPC pipleline using the $K$ reflected plane waves \cite{Bendjador2020}.  
From the generated DAS output tensor $\Zb$ in Fig.~\ref{fig:system_block_diagram}(a), Hilbert transformed
IQ data cube $\tilde \Zb$ is generated.  Then, we extract the local compound tensor $\Tilde{\Zb}_{ROI} \in \Rd^{A_r\times L_r\times K} $ of the region of interest (ROI), where $L_r$ and $A_r$ are the number of lateral and axial samples within a the ROI, respectively, and $K$ is the number of planewave measurements. The tensor $\Tilde{\Zb}_{ROI}$ is reshaped into a 2D Casorati matrix form as shown in Fig.~\ref{fig:system_block_diagram_SVD}.
The SVD of each $\Tilde{\Zb}_{ROI}$ matrix is then performed and provides the separation of spatial and angular variables. As described in \cite{Bendjador2020}, the first eigenvector guarantees the maximization of the angular coherence.

In contrast to previous studies \cite{robert2008green,vignon2017adaptive,nguyen2018spatial} where SVD is applied on subaperture or synthetic transmit data, this SVD beamformer \cite{Bendjador2020} merges Phase Aberration Correction (PAC) techniques and coherence-based imaging approaches in a unique matrix formalism implementation. 

\subsection{Proposed Method}
To deal with the limited measurement and phase aberration artifacts in planewave compounding, we replace the conventional CPC in Fig.~\ref{fig:system_block_diagram}(b) with the proposed 3D convolution neural network (3D CNN). 
As emphasized before, in contrast to the existing deep learning approaches that try to estimate the SoS distribution, we are interested in developing a phase aberration robust beamforming method  even without estimating the SoS distribution.  In particular, we implement a 3D CNN such that the neural network is trained to estimate artefact free image from $K$ number of planewave images generated with variable SoS in each line and each planewave. 
More details are as follows.

\subsubsection{Self-Supervised Training}
\begin{figure}[ht]
	\centerline{\includegraphics*[width=0.5\textwidth]{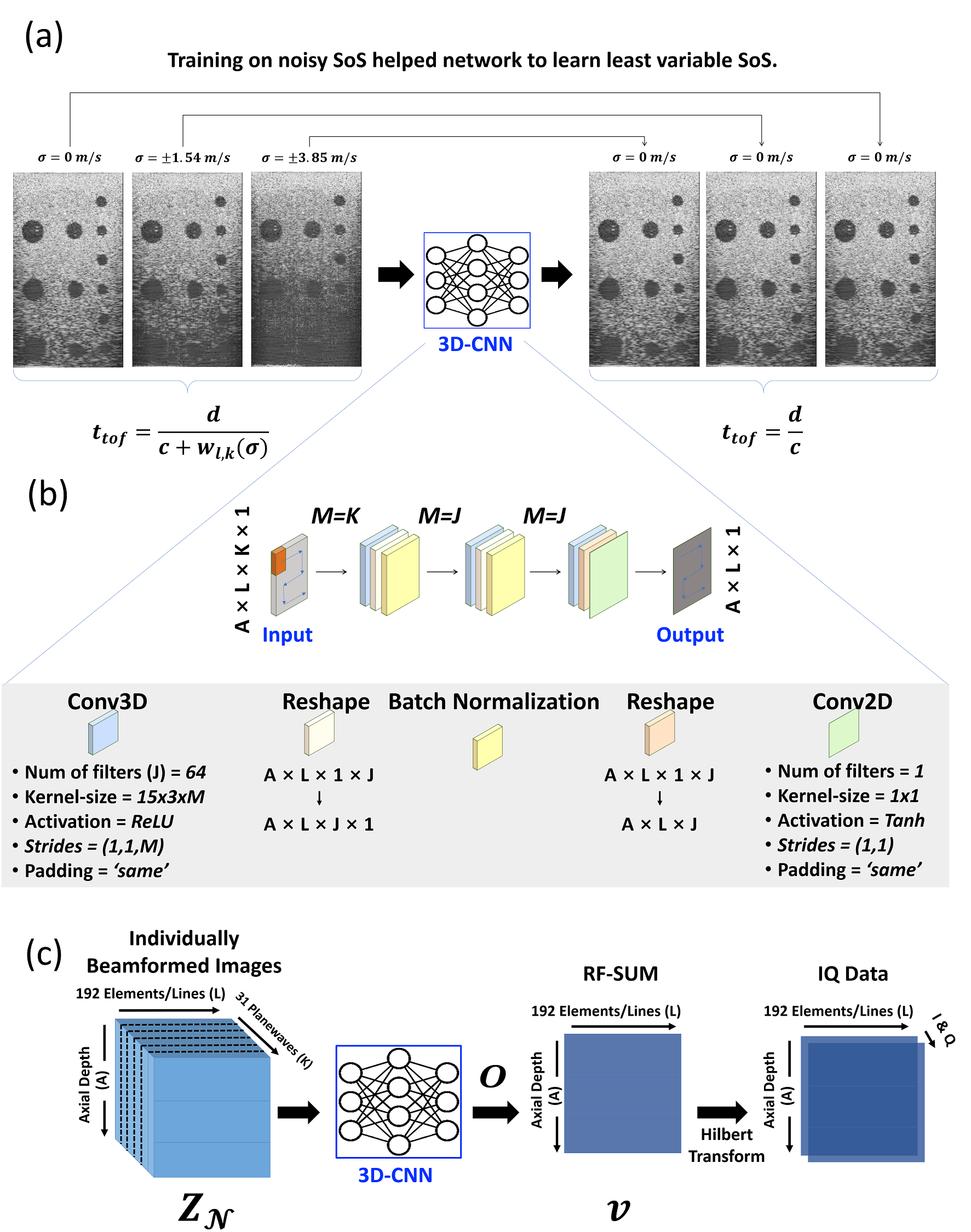}}
	\caption{Proposed phase aberration correction in planewave ultrasound using 3D-CNN.  (a) Proposed self-supervised training strategy. The network input is a noisy DAS output matrix $\Zb_{\Nc}$ generated using noisy SoS.  The training target is a CPC image $\vb$ generated using SoS of $c=1540m/s$.  (b) Architecture of the proposed 3D convolutional neural network. (a) Proposed data processing pipeline for robust planewave compounding. }
	\label{fig:architecture}	
\end{figure}

Specifically, the input of the proposed 3D-CNN is a noisy DAS tensor $\Zb_{\Nc}$ generated using noisy SoS $\tilde c_{l,k}$, which  is dependent upon
the lateral axis and angle direction:
$$\tilde c_{l,k} = c + w_{l,k}(\sigma),\quad w_{i,k}(\sigma)\sim U(-\sigma,\sigma),$$
 where $U(-\sigma,\sigma)$ is a uniform distribution between $-\sigma$ and $\sigma$.
 Then, the target  for the training is a CPC image $\vb$ generated using the standard value of SoS of $c=1540m/s$. 
 Fig.~\ref{fig:architecture}(a) shows the concept diagram of the proposed self-supervised training scheme. In order to estimate artifact-free image, the model is trained to produce the same image $\vb$ for three levels of variations $\sigma$ in SoS. In particular, we choose $\sigma \in \{0, 1.54, 3.85\}m/s$. This is based on our assumption that SoS variation can be viewed as random noise around the commonly used SoS value of $c=1540m/s$.

\subsubsection{Network Architecture}
The schematic diagram of proposed model is shown in Fig.~\ref{fig:architecture}(b). The proposed deep learning model consists of only three blocks, each block is composed of a three-dimensional (3D) convolution (Conv3D) layer and permutation/reshaping layers. In the first block, the input tensor $\Zb$ of size $A\times L\times K$  is first processed by a Conv3D layer consisted of $J=64$ number of 3D convolution filters, each of which has  size of $15\times3\times M$ where $M=K$ and \textit{Relu} activation function.  To reduce the computational complexity motivated by the deep block transform method \cite{jin2021deep}, the strides of the Conv3D layer were set to $(1,1,M)$. Later the output of the Conv3D layer is batch-normalized and permuted to reshape in $A\times L\times J$ format so that each output channel can be mixed with axial and lateral signals for next layer 3D convolution. All three convolution blocks have same configurations, except for the last block where the normalization step is removed and permutation/reshaping is followed by squeezing operation which reduces the dimensions so that a two-dimensional (2D) convolution (Conv2D) can be applied to merge the $J$ channels to a single planewave image using a filter of kernel-size $1\times1$. To maintain the output signal scale, at the last layer \textit{tanh} activation function is used which allows both positive and negative values.

Fig.~\ref{fig:architecture}(c) illustrate the proposed data processing pipeline for robust planewave compounding. Conceptually, the idea of the proposed phase aberration robust beamforming seems similar to the SVD-based beamforming. However, in contrast to the SVD-based beamforming, our method is computationally  efficient for real-time processing. Furthermore, unlike SVD beamformer \cite{Bendjador2020}, where IQ data cube is used as input (see Fig.~\ref{fig:system_block_diagram_SVD}), in the proposed data processing pipeline, DAS output tensor $\Zb$ is used as input signal for further reduction of the computational time.

Besides the computational advantage,  the most important advantage of our method is the self-supervised learning step. Specifically, unlike SVD-beamformer which cannot consider the variation of the SoS for the training,
our network is trained with various SoS variations so that the network becomes robust to such variations in the assumed SoS value.

\subsubsection{Loss function}

As for loss function, we simply use the $l_2$ loss function. Specifically, 
the  neural network training is given by
 \begin{eqnarray}\label{eq:train_new}
\min_{\Thetab} \sum_{t=1}^T\|\vb^{(i)} - \Tc_\Thetab\left(\Zb_{\Nc}^{(i)}\right) \|_2^2,
\end{eqnarray}
where $\Tc_\Thetab$ is a neural network parameterized by $\Thetab$, and 
 $\{(\Zb_\Nc^{(i)},\vb^{(i)})\}_{i=1}^{T}$ denotes the training data set composed of noisy DAS tensor 
 and the target CPC image with the standard SOS value.

{
 By minimizing the loss in \eqref{eq:train_new}, the proposed model can improve the quality of ultrasound images that have phase aberration artifacts, and furthermore, it improves the quality of input image when number of input planewaves are lower than the normal case.
}

\section{Method}
\label{sec:methods}

\subsection{Dataset}
 This research was carried out following the principles of the Declaration of Helsinki. In this study, we used real measurement data, which were acquired using an E-CUBE 12R US system (Alpinion Co., Korea).  For data acquisition, we used a linear array (L3-12H) transducer, and the operating configuration of probe is provided in Table \ref{probe_config}. 
\begin{table}[!hbt]
	\centering
	\caption{Linear Array Probe Configuration}
	\label{probe_config}
	\resizebox{0.3\textwidth}{!}{
		\begin{tabular}{c|c}
			\hline
			{Parameter} & {Linear array}  \\ \hline\hline
			Probe Model No. & L3-12H \\
			Carrier wave frequency & 8.48MHz \\
			Sampling frequency & 40 MHz \\
			Scan wave mode & 31-Planewaves \\
			No. of probe elements & 192 \\
			Elements pitch & 0.2 mm \\
			Elements width & 0.14 mm\\
			Elevating length & 4.5 mm \\
			Axial depth range & 20$\sim$80 mm \\
			Lateral length & 38.4 mm  \\
			Focal depth range & 10$\sim$40 mm \\\hline
	\end{tabular}}
\end{table}

\begin{figure}[ht]
	\centerline{\includegraphics[width=0.27\textwidth]{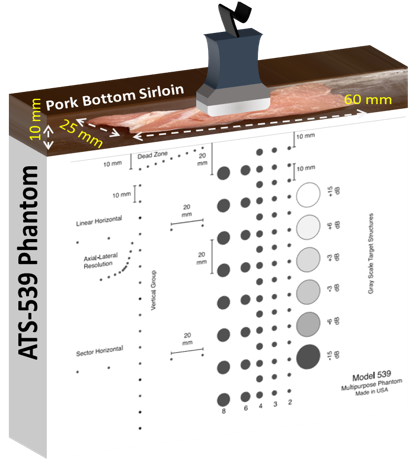}}
	\vspace*{-0.25cm}
	\caption{Schematic of the 10mm thick pork phantom. A Pork Bottom Sirloin cut of size $10\times25\times60$ is placed on ATS-539 phantom to produce phase abberation artifacts in tissue mimicking phantom scans.}
	\label{fig:pork_phantom_schematic}	
\end{figure}

In particular, we used the L3-12 probe operating at center frequency of $8.48$ MHz in a planewave mode. There are total $199$ scans acquired, $77$ from ATS-539 phantom and $122$ from \textit{in vivo} carotid/thyroid area of $4$ volunteers.  The dataset is processed using different subsets of plane waves (PWs) to simulate different acceleration factors.  During image formation,  scanline-wise variable speed-of-sound (SoS) is used to generate artificial phase aberration. In particular, three input subsets were generated for $31$PWs, $25$PWs, and $15$PWs with three SoS variation rates i.e., $\sigma \in \{0, 1.54, 3.85\}m/s$. The target images are generated using $31$ PW images processed using standard time-of-flight correction method of DAS/CPC method \cite{montaldo2009coherent}. Additionally, we acquired RF data from Pork Bottom Sirloin cuts of two different thickness placed on the ATS-539 phantom, which imposes real phase aberration from the different SoS from the fat. In particular, we used pork meat of 10 and 20 mm thickness. The schematic of the 10mm thick hybrid phantom is shown in Fig.~\ref{fig:pork_phantom_schematic}. 

\subsection{Comparison algorithm}
For comparative studies,  we compared our proposed method with standard CPC \cite{montaldo2009coherent} and a recently proposed SVD-based  aberration correction method \cite{Bendjador2020}. The SVD was implemented to process 31, 25 and 15 planewave images, with the patch-size of $32\times32$ (axial $\times$ lateral) samples. Finally, the contrast and computational cost of all three methods is compared for their Keras TensorFlow-based GPU implementation using Python.

\subsection{Performance metrics}
In this study, the standard quality metrics of ultrasound imaging including contrast statistics and computation time are used for the quantitative evaluation of the proposed method.
The contrast is measured in terms of contrast-recovery (CR), contrast-to-noise ratio (CNR), and generalized CNR (GCNR) \cite{rodriguez2019generalized} which are defined as follows:
\begin{equation}
{\hbox{CR}}(R_a,R_b) = |\mu_{R_a}-\mu_{R_b}|,
\end{equation}


\begin{equation}
{\hbox{CNR}}(R_a,R_b) = \frac{|\mu_{R_a}-\mu_{R_b}|}{\sqrt{\sigma^2_{R_a} + \sigma^2_{R_b}}},
\end{equation}
and
\begin{equation}
{\hbox{GCNR}}(R_a,R_b) = 1- \int \min \{p_{R_a} (i), p_{R_b} (i) \} di,
\end{equation}
where $i$ is the pixel intensity, $\mu_{R_a}$ and $\mu_{R_b}$ are the local means, $\sigma_{R_a}$ and $\sigma_{R_b}$ are the standard deviations, and $p_{R_a}$ and $p_{R_b}$ are the probability distributions of region ($R_a$) and ($R_b$), respectively.  Unlike CR and CNR, the range of GCNR score is bounded between zero and one. For the worst-case when two ROIs have identical intensity values, the GCNR will be equal to zero whereas in ideal case the GCNR must be equal to one.


\subsection{Network training}
For training, validation and testing, we used 8-fold cross validation protocol. In particular,  the dataset is first divided into 8 equal sized subsets. Then in each trial, 7 sub-sets were used for development (including 67\% training and 33\% validation) and a remaining subset was used for testing. The process is repeated 8 times and in each trial a different subset was used for testing. The training was performed to minimize the mean-squared error loss using Adam optimizer with an initial learning rate of $1\times 10^{-4}$ and it decreases exponentially with the decay rate of $1\times10^{-3}$  in each epoch. The batch-size of 5-samples was used and the model was trained for 300 epochs with the early {stopping patience of 5 epochs.}

\section{Results }
\label{sec:results}

\subsection{Normal and Artificial Phase Aberration Setting}

\subsubsection{Qualitative Analysis.}
For qualitative analysis we used one phantom and an \textit{in vivo} scan acquired from a carotid region of a healthy volunteer.  All B-mode images are shown for 60 dB dynamic range and the residual errors (the difference between CPC and the proposed method) are shown with pseudo-color in the  normalized range of $0\sim1$.

\begin{figure*}[ht]
	\centerline{\includegraphics[width=0.7\textwidth]{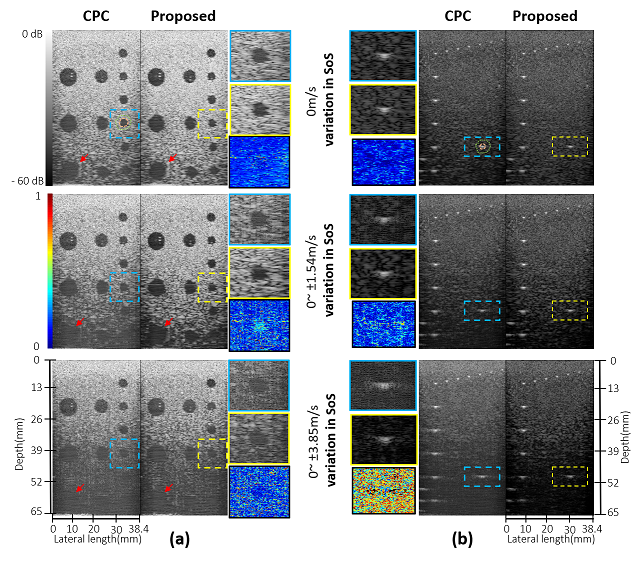}}
	\vspace*{-0.5cm}
	\caption{Comparison of standard coherent planewave compounding (CPC) and the proposed method for three different phase aberration settings using two phantom scans: (a) tissue mimicking phantom, (b) point target phantom.  All B-mode images are shown for 60 dB dynamic range and the residual-error (difference of CPC and proposed) are shown with pseudo-color in the normalized ranges of $0\sim1$.}
	\label{fig:Bmode_phantom}	
\end{figure*} 
Fig.~\ref{fig:Bmode_phantom}(a) shows a comparison of standard coherent planewave compounding (CPC) and the proposed method for three different artificial phase aberration settings. From the results, it can be easily seen that the proposed method noticeably improves the speckle details in both  near and far fields. Since the signal-to-noise ratio (SNR) of the acoustic signal drops with the axial depth, we can clearly see that in the deep (far) regions it is becoming difficult to discriminate between the background and the hypo-echoic phantom (see red arrows in Fig.~\ref{fig:Bmode_phantom}(a)). Interestingly, the proposed method showed resistance to the drop in SNR for a wide range of axial depth and variations in the speed of sound. 

An interesting feature of the proposed method is that a single one-time trained model can be used for different levels of phase aberrations. This is important in real world application as characterization of noise/artefact levels is a challenging task, and from the image alone it is challenging to decide what level of SoS correction is needed.  It is also noteworthy to point-out that the proposed model is trained on simulated random variations in SoS using noisy samples without ideal (artefact-free) ground truth. However, it show noticeable gain in structural details for $0\sim \pm1.54$ m/s variations, despite the fact that the random variations are performed in each scan line and each plane wave.  
Under no artificial variations in SoS i.e., 0 m/s variation case,  the output of proposed method is still noticeably improved. This  may be because even when there is no simulated phase aberration, the proposed model automatically improve the image quality and correct for natural phased aberration caused during the measurement.

To further evaluate the robustness of the proposed model, in Fig.~\ref{fig:Bmode_phantom}(b), the results on point target phantom are provided. The point target phantom is used to show the robustness in resolution preservation. On careful observation it is evident that the contrast and structural detail in the proposed method is maintained for all three levels of SoS variations. The normalized residual error of the selected regions are shown in side figures, where it can be clearly seen that with the increase in SoS variations, speckle patterns in CPC become too noisy and B-mode image show washout artefacts, which is one of the many characteristics of phase-aberration \cite{anderson2000impact}.

\begin{figure*}[ht]
	\centerline{\includegraphics[width=0.8\textwidth]{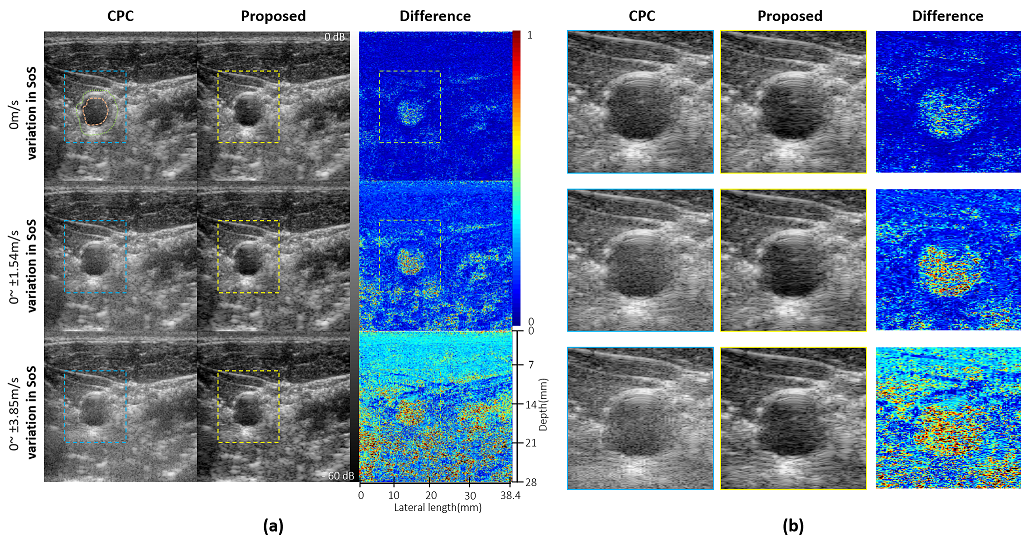}}
	\vspace*{-0.5cm}
	\caption{Comparison of standard coherent planewave compounding (CPC) and the proposed method for three different phase aberration settings using an \textit{in vivo} scans from carotid region of a healthy volunteer: (a) fully scan B-mode results by CPC, the proposed method, and their difference, (b)
	magnified view of the blue and yellow marked regions from full images in (a).  All B-mode images are shown for 60 dB dynamic range and the residual error (difference of CPC and proposed) are shown with the pseudo-color at the  normalized ranges of $0\sim1$.}
	\label{fig:Bmode_invivo}	
\end{figure*} 

To evaluate the utility of proposed method, we tested our method on \textit{in vivo} samples.  Fig.~\ref{fig:Bmode_invivo}(a) shows a comparison of standard coherent planewave compounding (CPC) and the proposed method for three different artificial phase aberration settings using an \textit{in vivo} scan from carotid region of a healthy volunteer. Since  the tissue characteristics in the case of \textit{in vivo} scan are not as smooth as in tissue mimicking phantom,  we can expect that there exist some SoS variations even when there is no artificial variations. Here again we can see that the proposed method has relatively high tolerance for the variations in SoS and therefore the washout artefacts are noticeably reduced.  To better visualize the quality enhancement in the proposed method, the blue and yellow marked regions from full images are zoomed and shown in Fig.~\ref{fig:Bmode_invivo}(b). In both  original and residual images, it is evident that the contrast and structural details are enhanced.

\begin{figure*}[ht]
	\centerline{\includegraphics[width=0.6\textwidth]{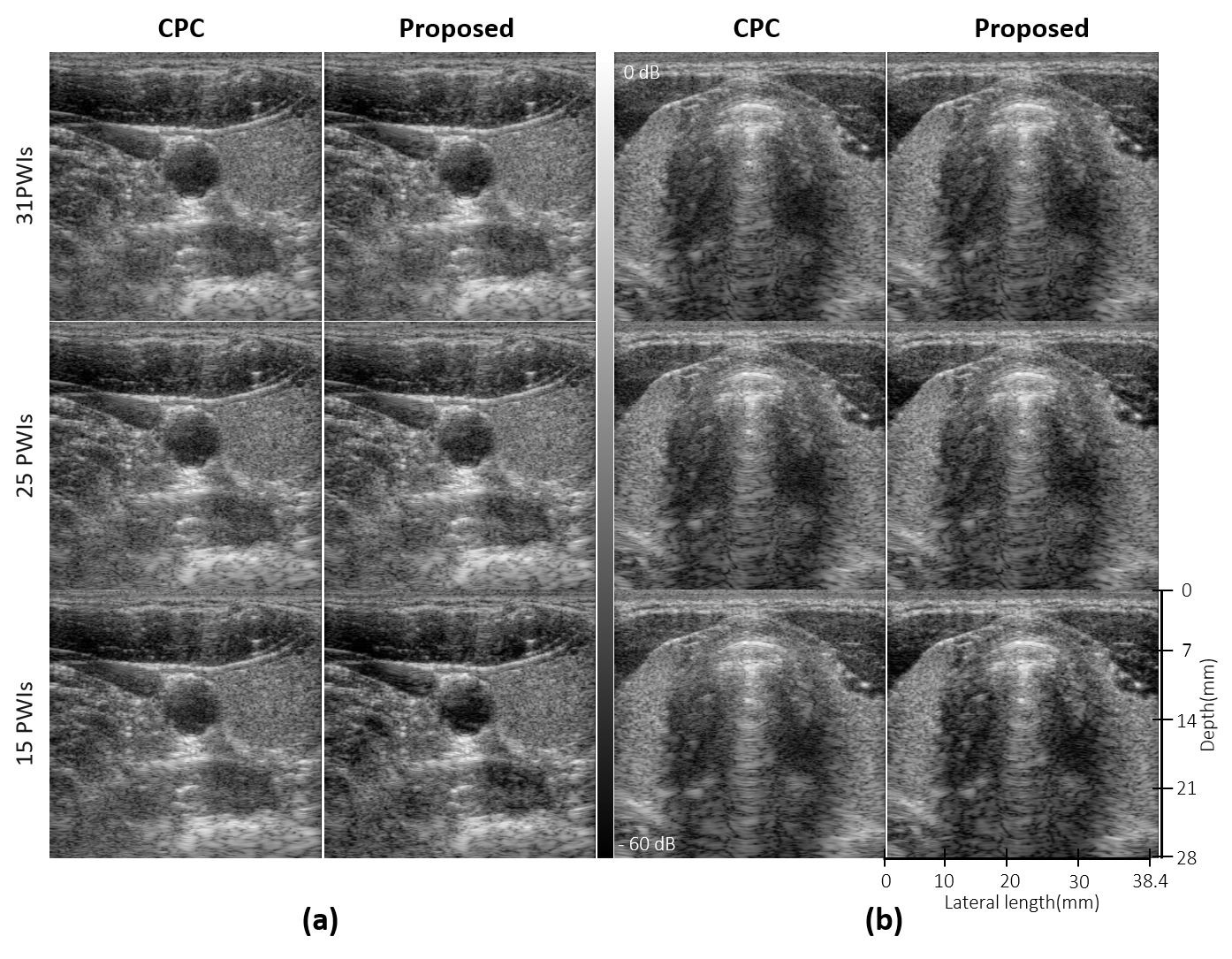}}
	\vspace*{-0.5cm}
	\caption{Comparison of standard coherent planewave compounding (CPC) and the proposed method for three different measurement settings using an \textit{in vivo} scans from (a) carotid and (b) trachea regions of a healthy volunteers: All B-mode images are shown for 60 dB dynamic range.}
	\label{fig:Bmode_invivo_extra}	
\end{figure*} 

To show the performance gain with reduced measurement case in Fig.~\ref{fig:Bmode_invivo_extra}, two additional examples are shown for three different numbers of planewave images. In both examples, it can be clearly seen that the proposed method has higher tolerance for measurement loss and the contrast and structural information is well retained.

\subsubsection{Quantitative Analysis}
\begin{figure*}[ht]
	\centerline{\includegraphics*[width=0.8\textwidth]{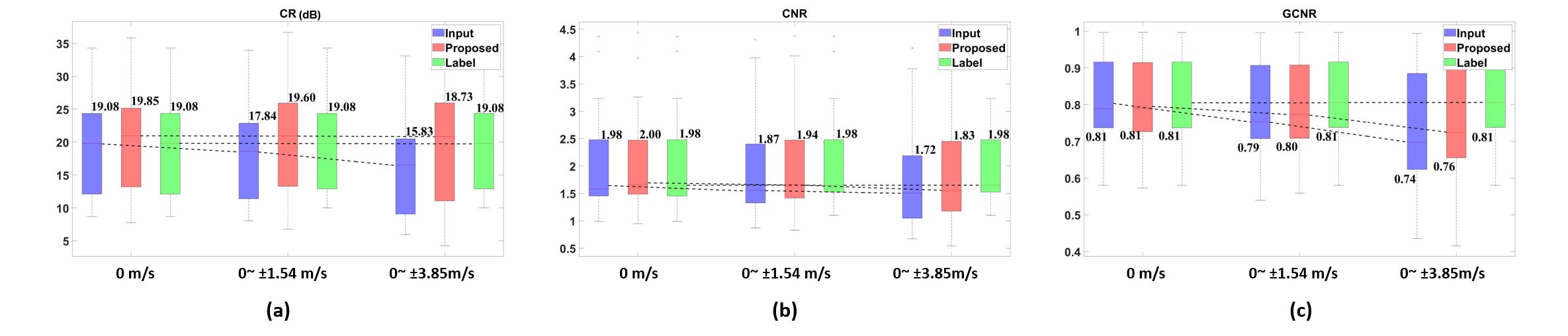}}
	\caption{8-fold cross validation performance statistics of  input, standard coherent planewave compounding (CPC), and the proposed method for three different phase aberration settings on complete dataset. (a) Contrast recovery (CR), (b) Contrast-to-noise ratio (CNR), and (c) Generalized contrast-to-noise ratio (GCNR).}
	\label{fig:contrast_stats}	
\end{figure*} 

To quantify the contrast gain of the proposed method, we compare the distributions of test statistics using contrast metrics defined in Section~\ref{sec:methods}. As anticipated, in Fig.~\ref{fig:contrast_stats}(a) the CR values of the proposed method shows consistent results for all three variations. In particular for $0$ m/s, $0\sim\pm1.54$ m/s, and $0\sim\pm3.85$ m/s variations, the proposed method recovered 0.77, 1.76 and 2.9 dB better contrast. Considering the vulnerability of CR metric on measurement noise, one may say that the contrast itself is not a suitable metric. To quantify contrast in the presence of noise, a standard metric is CNR. In CNR metric plots shown in Fig.~\ref{fig:contrast_stats}(b), the CNR values of the proposed method shows superior performance for all three SoS variations.  In particular, compared to CPC, there is 0.02, 0.07 and 0.11 units gain in CNR for $0$ m/s, $0\sim\pm1.54$ m/s, and $0\sim\pm3.85$ m/s SoS variations, respectively.

In recent past, both the CR and CNR metrics are criticized for their sensitivity over different pre-processing steps, e.g., variations in dynamic range, non-linear filtration and other cosmetic changes in signal intensities \cite{rodriguez2019generalized}. Therefore, for a fair comparison, we used GCNR statistics to see whether the contrast gain achieved through the proposed scheme is real or just a cosmetic change due to distribution shift in pixel intensities. Although in GCNR the scale of gain is small, but  the proposed method showed consistently improved performance and the results are 0.001, 0.010 and 0.020 units higher than the conventional CPC method for $0$ m/s, $0\sim\pm1.54$ m/s, and $0\sim\pm3.85$ m/s SoS variations, respectively. It is worthy to point out that the low performance gain in $0$ m/s case is understandable as there is no noise-free label data for training and the proposed model is trained to mimic fully-sampled 31-PW CPC results from three levels of SoS variations: therefore a well trained model should maintain the CPC like results. 

\subsubsection{Comparative Analysis}

\begin{figure}[ht]
	\centerline{\includegraphics[width=0.4\textwidth]{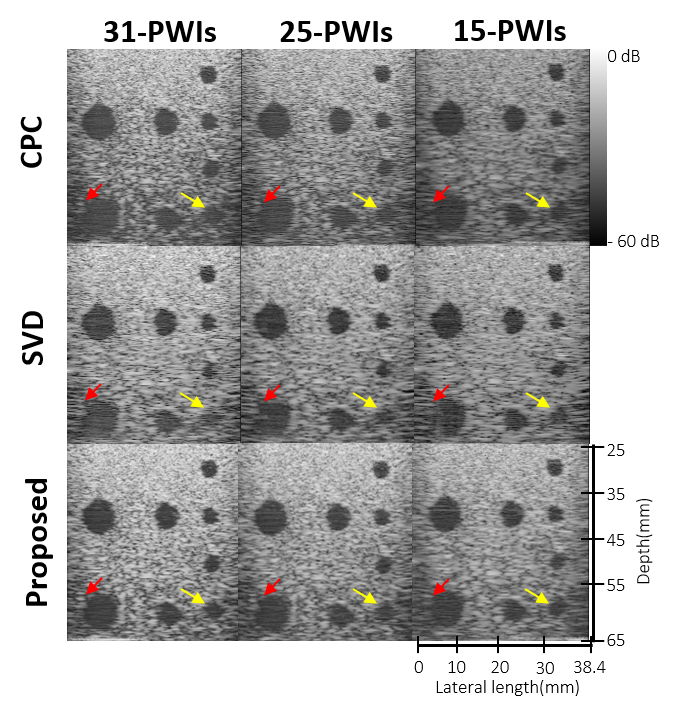}}
	\vspace*{-0.5cm}
	\caption{B-mode images of tissue mimicking phantom by standard coherent planewave compounding (CPC), SVD beamformer \cite{Bendjador2020} and the proposed method using 31, 25 and 15 planewave images (PWIs). All B-mode images are shown for 60 dB dynamic range.}
	\label{fig:BMobe_CPC_SVD_DeepCPC}	
\end{figure} 

\begin{figure*}[ht]
	\centerline{\includegraphics[width=0.8\textwidth]{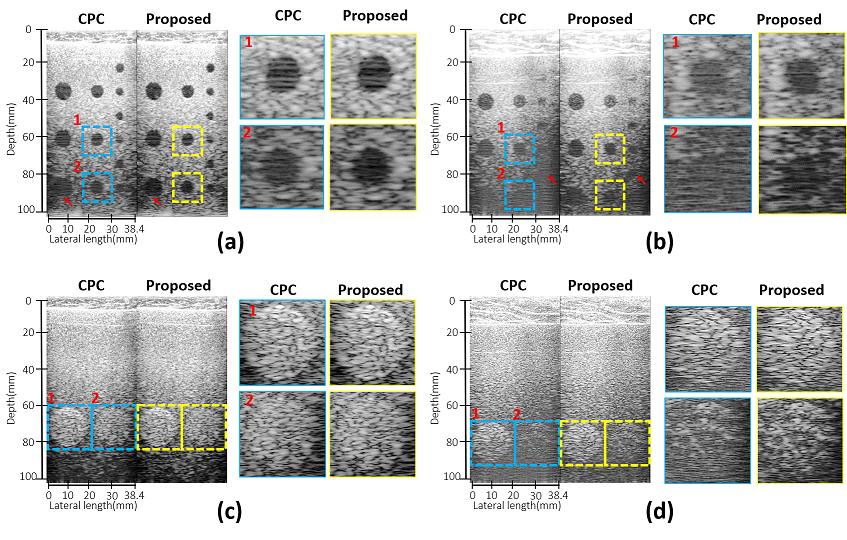}}
	\vspace*{-0.5cm}
	\caption{Comparison of standard coherent planewave compounding (CPC) and the proposed method using pork phantom for two different thickness to produce phase aberration in ATS-539 scans: (a) 10mm thick pork lion cut on a hypo-echoic phantom, (b) 20mm thick pork lion cut on a hypo-echoic phantom, (c) 10mm thick pork lion cut on a hyper-echoic phantom, (d) 20mm thick pork lion cut on a hyper-echoic phantom. The magnified view of two regions is shown as side figures where blue and yellow bordered figures represent the CPC and proposed methods, respectively. All B-mode images are shown for 60 dB dynamic range.}
	\label{fig:BMobe_pork}	
\end{figure*}

For comparative analysis, we compare our method with a recently proposed SVD beamformer (SVD-BF) \cite{Bendjador2020}. 
As explained before, the SVD-BF is a robust beamforming technique for phase aberration correction. 
From the results in Fig.~\ref{fig:BMobe_CPC_SVD_DeepCPC}, it can be seen that the contrast and structural details in SVD-based planewave compounding are noticeably better than the standard CPC method. However, its performance is relatively lower compared to the proposed method which works for both the deep feature recovery and maintain superior performance even with only 15 PWIs (see red and yellow arrows). This is important because in planewave imaging, the frame-rate of scan is determined by the number of planewaves. 

\begin{table*}[ht]
	\centering
	\caption{Comparison of CPC, SVD and the proposed method.}
	\label{tbl:results_PWI_enhacement}
	\resizebox{16cm}{!}{
		\begin{tabular}{c|ccc|ccc|ccc}
			\hline
			number of & \multicolumn{3}{c|}{CR (dB)} & \multicolumn{3}{c|}{{CNR}} & \multicolumn{3}{c}{{GCNR}}  \\
			{PWs} & \textit{a} &  \textit{b} &  \textit{c} & \textit{a} &  \textit{b} &  \textit{c} & \textit{a} &  \textit{b} &  \textit{c} \\ \hline\hline
			15 & 19.00$\pm$5.62 & 18.87$\pm$5.40 & 19.85$\pm$6.21 & 1.96$\pm$0.61 &	1.95$\pm$0.58 & 2.00$\pm$0.63 & 0.809$\pm$0.103 & 0.810$\pm$0.099 & 0.814$\pm$0.103\\
			25 & 19.2$\pm$5.95 & 19.16$\pm$5.68 & 20.31$\pm$6.22 & 1.98$\pm$0.64 & 1.98$\pm$0.60 & 2.02$\pm$0.63 & 	0.814$\pm$0.102 & 0.813$\pm$0.099 & 0.816$\pm$0.102 \\
			31 & 19.08$\pm$6.15 & 19.25$\pm$5.72 & 20.77$\pm$6.36 & 1.98$\pm$0.65 & 1.99$\pm$0.60 & 2.03$\pm$0.66 & 0.815$\pm$0.103 & 0.816$\pm$0.097 & 0.816$\pm$0.102 \\  \hline
		\end{tabular}
	}
	\\
	\small{$^a$ CPC, $^b$ SVD, $^c$ Proposed}
	\vspace{0cm}
\end{table*}

To quantify the performance gain achieved with the proposed method, Table~\ref{tbl:results_PWI_enhacement} compared contrast statistics of CPC, SVD and the proposed method for 31, 25 and 15 PWIs. Here again we can clearly see that the proposed method can work for reduced number of planewave images to provide similar performance. Hence the method is robust against measurement loss as well as SoS variations. With $2.07\times$ acceleration i.e., using only 15 planewaves, the proposed method can recover significantly (i.e., p-value $<10^{-3}$) improved results. In particular, $0.85$, and $0.98$ decibels higher CR, $0.04$, and $0.05$ units higher CNR, and $0.05$, and $0.04$ units higher GCNR are achieved compared to standard CPC and SVD methods, respectively. 

Furthermore, the computational complexity of the proposed method is also quite comparable to that of the CPC, which is another added advantage over SVD-BF. In particular, for a 17.5mm deep and 38.4mm wide scan consisting of $1024 \times 192$ (axial $\times$ lateral) samples, the CPC method takes 9.28, 7.20 and 4.75 mSecs for 31, 25 and 15 PWIs, respectively. Similarly, for the same sized image using $32\times 32$ patch processing, the SVD method takes 23012.80, 22795.46 and 22559.35 mSecs for 31, 25 and 15 PWIs, respectively, which is several magnitude higher than the proposed method which takes 70.64, 66.72 and 47.74 mSecs for 31, 25 and 15 PWIs, respectively.

\subsection{Self-induced Real Phase Aberration using Pork Phantom}
To produce phase aberration artifacts in real measurements, we design an experiment with the pork phantom. In particular four scans were acquired using 10 and 20 mm thick Pork Bottom Sirloin cuts placed on the ATS-539 phantom. The schematics of a 10 mm thick hybrid phantom is shown in Fig.~\ref{fig:pork_phantom_schematic}.

Fig.~\ref{fig:BMobe_pork} show the reconstruction results of the standard CPC and the proposed method for the pork phantom. From the results it can be clearly seen that the proposed method improves the overall quality of the images, especially in the scans with 20mm thick pork cut (see region 2 in Fig.~\ref{fig:BMobe_pork}(a) and (b)) where mid-size hypo-echoic phantom is virtually invisible in CPC image. A similar effect scan be seen in hyper-echoic scans (see region 2 in Fig.~\ref{fig:BMobe_pork}(c) and (d)) where the image of the phantom in region 2 is completely distorted. On careful observation it can also be seen that the proposed method noticeably improves the contrast and structural details in all four cases, especially in deeper scans (see red arrow in Fig.~\ref{fig:BMobe_pork}(a) and (b)). This confirms that the proposed method which is trained in a self-supervised fashion can reduce the phase aberration artifacts in real measurement and improve the overall quality of the B-mode imaging.

\section{Conclusion}
\label{sec:conclusion}
In this  paper, we presented a purely data-driven method for the compounding of planewave ultrasound images.  The proposed method was trained in a self-supervised manner on simulated phase aberration artifacts and tested on three different SoS variations and acceleration factors. 
Moreover, the network architecture was intelligently designed to reduce computational overhead and its computationally complexity is comparable to the standard CPC and SVD based methods. 
Experimental results showed that the proposed neural network-based reconstruction approach is robust and suitable for deep scans where SoS variations create most of the artefacts. The proposed model was shown to achieve significant improvement for both the {phantom} and the {in vivo} scans. 


\end{document}